\documentclass[twocolumn,aps,prl,superscriptaddress,longbibliography]{revtex4-2}

\pdfoutput=1

\usepackage{amssymb}
\usepackage{amsmath}
\usepackage{graphicx,bm}
\usepackage{epstopdf}
\usepackage{subfigure}
\usepackage{natbib}
\usepackage{epsfig}
\usepackage{amsfonts}
\usepackage{mathrsfs}
\usepackage{CJK}
\usepackage[toc,page,title,titletoc,header]{appendix}
\usepackage{array,longtable}
\usepackage{float}
\usepackage[usenames, dvipsnames, x11names]{xcolor}
\usepackage[colorlinks]{hyperref}
\usepackage[capitalise]{cleveref}
\usepackage{enumerate}
\usepackage{soul}
\usepackage{dsfont}
\hypersetup{
 colorlinks=true,
 citecolor=red,
 linkcolor=blue,
 urlcolor=cyan}
\usepackage{mathtools}

\newcommand{\R}{\mathbb{R}}

\renewcommand{\H}{\hat{\mathcal{H}}}
\newcommand{\D}{\mathcal{D}}

\newcommand{\0}{\overline{0}}
\newcommand{\1}{\overline{1}}

\usepackage{comment}

\begin{document}
\title{Dissipative Pairing Interactions: Quantum Instabilities, Topological Light, and Volume-Law Entanglement}

\author{Andrew Pocklington}
\affiliation{Pritzker School of Molecular Engineering, University of Chicago, 5640 South Ellis Avenue, Chicago, Illinois 60637, USA}

\affiliation{Department of Physics, University of Chicago, 5640 South Ellis Avenue, Chicago, Illinois 60637, USA}

\author{Yu-Xin Wang}
\affiliation{Pritzker School of Molecular Engineering, University of Chicago, 5640 South Ellis Avenue, Chicago, Illinois 60637, USA}

\author{A. A. Clerk}
\affiliation{Pritzker School of Molecular Engineering, University of Chicago, 5640 South Ellis Avenue, Chicago, Illinois 60637, USA}

\begin{abstract}
We analyze an unusual class of bosonic dynamical instabilities 
that arise from dissipative (or non-Hermitian) pairing interactions.
We show that, surprisingly, a completely stable dissipative pairing interaction can be combined with simple hopping or beam-splitter interactions (also stable) to generate instabilities. Further, we find that the dissipative steady state in such a situation remains completely pure up until the instability threshold (in clear distinction from standard parametric instabilities). 
These pairing-induced instabilities also exhibit an extremely pronounced sensitivity to wave function localization. This provides a simple yet powerful method for selectively populating and entangling edge modes of photonic (or more general bosonic) lattices having a topological band structure. The underlying dissipative pairing interaction is experimentally resource-friendly, requiring the addition of a single additional localized interaction to an existing lattice, and is compatible with a number of existing platforms, including superconducting circuits.
\end{abstract}

\maketitle


{\textit{Introduction.---}}Hamiltonian bosonic pairing interactions (where excitations are coherently created or destroyed in pairs) arise in many settings, and underpin a vast range of phenomena. In the context of quantum optics and information, they are known as parametric amplifier interactions, and are a basic resource for generating squeezing and entanglement \cite{Caves1985,Gerry1985}; they also form the basis of quantum limited amplifiers \cite{Louisell1961,Metelmann2014}. In condensed matter settings, bosonic pairing underlies the theory of antiferromagnetic spin waves, interacting Bose condensates, and can also be used to realize novel topological band structures \cite{Shindou2013,Peano2016}.

Given the importance of bosonic pairing, it is interesting to explore the basics of purely {\it dissipative} (or non-Hermitian) bosonic pairing. Non-Hermitian dynamics have garnered attention in a wide range of fields, from condensed matter \cite{Bergholtz2021,McDonald2022,Lieu2018} to optics \cite{El-Ganainy2019,Chen2017,Hodaei2017} to classical dynamical systems \cite{Fruchart2021,Scheibner2020,Sone2020}. 
In this Letter, we provide a comprehensive analysis of dissipative bosonic pairing in a fully quantum setting, showing it possesses a number of surprising and potentially useful features. We focus on minimal, experimentally realizable models, where bosons (e.g.~photons) hop on a lattice, in the presence of a single dissipative pairing interaction. Remarkably, we find that while the dissipative pairing interaction on its own yields fully stable dynamics, when combined with simple lattice hopping (which is also stable), one can have dynamical instability. Further, close to such an instability, the quantum steady state is perfectly pure, with a selected subset of modes having high densities and strong squeezing and/or entanglement correlations. The complete state purity up until the instability threshold is a clear distinction from more standard instabilities associated with Hermitian pairing terms. Dissipative pairing is also distinct from the well-studied situation where a system is driven with squeezed noise; in particular, driving a quadratic, particle-conserving system with squeezed noise can never generate instability, whereas this readily occurs with dissipative pairing.
\begin{figure}[t]
 \centering
 \includegraphics[width = 3.4in]{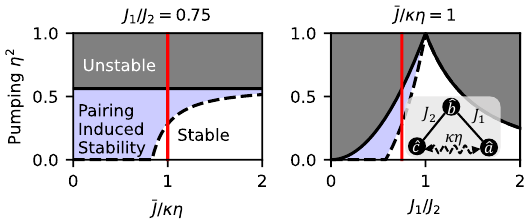}
 \caption{Stability diagram for a minimal three-mode bosonic system (see inset) with loss on mode $\hat{a}$ (rate $\kappa$), gain on mode $\hat{c}$ (rate $\eta^2 \kappa$), and tunnel couplings $J_1, J_2$ [cf. \cref{eq:DM}]. In the absence of dissipative pairing, the system is dynamically unstable above the dashed line. Adding dissipative pairing $i \eta \kappa (\hat a \hat c + \mathrm{H.c.})$ shifts the onset of instability to the solid line, see \cref{eqn:instability}. Remarkably, this boundary is independent of $\overline{J} / \kappa \eta$, where $\overline{J} = \sqrt{J_1^ 2+ J_2^2}$. The dissipative steady state remains pure (with a high density) as one approaches instability, see main text. Red lines in each plot are the same cut of parameter space, $\overline{J}/\kappa \eta = 1$ and $J_1/J_2 = 0.75$. Solid lines show hopping, dashed line shows the dissipative pairing interaction.} 
 \label{fig:FGR}
\end{figure}

Dissipative pairing becomes even more interesting when combined with topological band structures. We find that our new pairing instabilities are highly susceptible to wave function localization of the underlying lattice Hamiltonian. Hence, if the lattice supports exponentially localized topological edge modes, we are able to selectivity excite and entangle them. Such topological systems remain a cornerstone of condensed matter physics \cite{Qi2011,Kane2010,Thouless1982} and photonics \cite{Ota2020,Ozawa2019,Lu2014}, and selectively exciting edge modes has been the subject of a flurry of recent proposals \cite{Mittal2018,Barnett2013,Hu2021,Peano2018,Engelhardt2016,Secli2021}. These are motivated by applications including topological lasing \cite{Hu2021,Secli2021,StJean2017,Parto2018,Zhao2018} and topological amplification and squeezing \cite{Peano2018,Porras2019,Duenas2021,Wanjura2020}. However, these proposals often require complicated momentum and/or energy selectivity \cite{Peano2018,Engelhardt2016,Secli2021}, as well as control over the entire lattice, \cite{Peano2018,Engelhardt2016,Secli2021,Porras2019}. Here, we are able to get edge-mode selectivity almost for free, using a single quasilocal dissipative interaction.


{\textit{Minimal model.---}}We start with a three-mode system (bosonic annihilation operators $\hat{a},\hat{b},\hat{c}$) that exhibits much of the surprising physics of interest. The key ingredient will be a dissipative pairing interaction between $\hat a$ and $\hat c$, that is an interaction generating dynamics of the form $\partial_t \langle \hat a \rangle = - \lambda \langle \hat c ^\dagger \rangle$ and $\partial_t \langle \hat c^\dag \rangle =  \lambda^* \langle \hat a \rangle$. Because of the relative sign here, this dynamics {\it cannot} be obtained from a Hermitian pairing interaction. Instead, it would seem to correspond to a non-Hermitian effective Hamiltonian: 
\begin{align}
 \H_{\text{pairing}} = -i(\lambda \hat a^\dagger \hat c^\dagger + \mathrm{H.c.}). 
 \label{eqn:pairing}
\end{align}

To obtain this Markovian dissipative dynamics in a fully quantum setting, this dissipative interaction must necessarily be accompanied by noise as well as local damping and antidamping \cite{McDonald2022,Lau2018}. The resulting description has the form of a Lindblad master equation \cite{Lindblad1976,Gorini1976}. Using a minimal noise realization of the interaction, and letting $\hat{\rho}$ denote the system density matrix, we obtain
\begin{align}
 \dot{\hat \rho} &= \hat L \hat \rho \hat L^\dagger - \left\{ 
 \frac{\hat L^\dagger \hat L}{2}, \hat \rho \right\} \equiv \D [\hat L] \hat \rho,
 \,\,\,
 \hat L = \sqrt{\kappa} \hat a + \eta \sqrt{\kappa} \hat c ^\dagger.
 \label{eqn:master}
\end{align}
This purely dissipative evolution generates local damping on $\hat a$ with strength $\kappa$, local antidamping on $\hat c$ with strength $\eta^2 \kappa$, and a dissipative interaction of the form of Eq.~(\ref{eqn:pairing}) with $\lambda = \eta \kappa / 2$. We take $\eta < 1$ (i.e.~more local damping than antidamping), which ensures dynamical stability (i.e.,~no tendancy for exponential growth) \cite{Supplement}.

The dissipation in Eq.~(\ref{eqn:master}) is reminiscent of the dynamics generated by driving modes $\hat{a}$, $\hat{c}$ with broadband two-mode squeezed (TMS) noise \cite{Kraus2004}. There are, however, crucial differences. Driving with TMS noise always generates two dissipators; to make Eq.~(\ref{eqn:master}) equivalent to injected TMS nosie, we would thus have to add the additional dissipator $\mathcal{D}[\sqrt{\kappa} \hat c + \eta \sqrt{\kappa} \hat a^\dagger]$. This complementary dissipator would completely cancel the effective dissipative interaction between $a$ and $c$ generated by $\mathcal{D}[\hat{L}]$, leaving only driving with correlated noise. There would thus be no interaction from the dissipation in the equations of motion between $\langle \hat{a}(t) \rangle$ and $\langle \hat{c}^\dagger(t) \rangle$. In contrast, we will show that in \cref{eqn:master}, the direct dissipative interaction between modes $\hat a$ and $\hat c$ plays a crucial role. 

To see explicitly that dissipative pairing is distinct from input TMS noise, we will add coherent hopping interactions to our system, and consider the evolution of average values. The hopping is described by 
$ \H = J_1 \hat a^\dagger \hat b + J_2 \hat b^\dagger \hat c + \mathrm{H.c.}$,
with the evolution now given by $\partial_t \hat{\rho} = -i [ \H, \hat{\rho} ] + \mathcal{D}[\hat{L}] \hat{\rho}$. Because of linearity, the equations of motion for averages of mode operators are insensitive to noise, and only influenced by interactions (coherent and dissipative). For our system, a symmetry argument \cite{Supplement} lets us reduce the dynamics of these averages to the closed linear dynamics of the quadratures $\vec v = (x_a, p_b, x_c)$, where $\langle \hat a \rangle = (x_a + i p_a)/\sqrt{2}$, etc; the orthogonal quadratures $(p_a, x_b, p_c)$ have an analogous closed evolution. We find
$\partial_t \vec v = -iD \vec v$, where the dynamical matrix $D = D_J + D_\kappa$ can be interpreted as an effective $3 \times 3$ Hamiltonian matrix, and
\begin{align}
D_J &= \left(
\begin{array}{ccc}
0 & i J_1 & 0 \\
- i J_1 & 0 & - i J_2 \\
 0 & i J_2 & 0
\end{array} 
\right),
\,\,
D_\kappa = \frac{\kappa}{2} \left(
\begin{array}{ccc}
-i & 0 & -i \eta \\
0 & 0 & 0 \\
i \eta & 0 & i \eta^2
\end{array} 
\right). \label{eq:DM}
\end{align}
The off-diagonal $\pm i \eta \frac{\kappa}{2}$ terms in $D_\kappa$ are the dissipative interaction, which surprisingly adds a \textit{Hermitian} contribution at the level of the dynamical matrix. This mirrors the fact that had we started with a nondissipative Hermitian pairing interaction, we would generate a \textit{non-Hermitian} dynamical matrix \cite{Wang2019}. Note that the hopping dynamics on its own generates stable dynamics, as does the dissipative dynamics on its own. More formally, both the matrices $D_J$ and $D_\kappa$ have no eigenvalues with positive imaginary part and hence are dynamically stable (in the Lyapunov sense \cite{Hirsch2012} that there is no tendency for exponential growth).

We now come to our first surprise: while each part of our dynamics (hopping, dissipation) is stable individually, combining them can lead to instability. We find that for the full dynamics, whenever $J_1 \neq J_2$, there will be a critical value of $\eta$ beyond which we have exponential growth. Specifically, one can show \cite{Supplement} that the dynamical matrix in \cref{eq:DM} will be unstable if
\begin{align}
 \eta > \min\left( \left| J_1/J_2 \right|, \left|J_2/J_1 \right| \right).  \label{eqn:instability}
\end{align}
We stress that this phenomenon is distinct from recently studied ``dissipation-induced instabilities'' \cite{Dogra2019}, where the purely dissipative dynamics is already unstable on its own. Again, in our case the system is always stable in the dissipation-only limit $J_1 = J_2 = 0$. 

The instability threshold \cref{eqn:instability} can be understood from a simple perturbative argument that is formally valid only when $\kappa \ll J_1, J_2$ [akin to a Fermi's golden rule (FGR) calculation]. If we define $|\psi_i\rangle (i = 1,2,3)$ to be the (nondegenerate) eigenvectors of $D_J$, and treat $D_\kappa$ as a small perturbation on top of this, then to first order $|\psi_i\rangle$ has a relaxation rate
\begin{align}
 \Gamma_i &= -{\rm Im} \langle \psi_i | D_\kappa | \psi_i \rangle. \label{eqn:FGR}
\end{align}
If an eigenmode has more amplitude on $\hat c$ than $\hat a$, there will be a value of $\eta < 1$ at which \cref{eqn:FGR} is negative. This corresponds exactly to the condition in \cref{eqn:instability}, and is easy to understand intuitively (i.e.~the eigenmode sees more antidamping than damping). Surprisingly, this simple FGR argument turns out to be exact to all orders in $\kappa$: \cref{eqn:instability} is not perturbative \cite{Supplement}. We stress that this is a nonobvious phenonmenon. For example, consider a modified model where we eliminate dissipative pairing by replacing $\mathcal{D}[\hat{L}] \rightarrow \mathcal{D}[\sqrt{\kappa}\hat{a}] + 
\mathcal{D}[\sqrt{\kappa} \eta \hat{c}^\dagger]$ in our master equation. We are left with just incoherent gain and loss. In this case, the instability threshold would depend sensitively on the value of $\kappa$, with the FGR prediction only valid for $\kappa \rightarrow 0$, see \cref{fig:FGR}.

We thus see that even at the semiclassical level, the dissipative pairing interaction yields surprises: instability from the combination of two individually stable dynamical processes, with a threshold that is independent of the overall dissipation scale. Note that the above phenomena could alternatively be described (in a squeezed frame) as the interplay of asymmetric loss and Hermitian pairing interacting (see \cite{Supplement} for details and application to two-mode models). 

{\textit{Extension to quantum lattices.---}}We now explore dissipative pairing in general multimode lattice systems, focusing on the possibility of nontrivial dissipative steady states. Consider an $N$-site bosonic lattice, with annihilation operators $\hat a_i$ for each site. The coherent dynamics corresponds to a quadratic, number conserving Hamiltonian $\H = \sum_{ij} H_{ij} \hat a_i^\dagger \hat a_j$. The only constraint we impose is that $H$ possesses an involutory chiral sublattice symmetry $U$, such that $U H U^{\dagger} = -H$; our simple three-site model also had this symmetry. Chiral symmetry ensures that for every eigenmode of $H$ with nonzero energy, there is a different eigenmode with an opposite energy. 

We now add a single dissipative pairing interaction to the lattice, between two arbitrary sites $\0,\1$. Motivated by our three-mode example, we take $\0,\1$ to be on the same sublattice (as defined by the chiral symmetry). The full dynamics on the lattice is  given by \cite{PRBL} \nocite{Pocklington2022}
\begin{align}
 \partial_t \hat{\rho} = -i [ \H, \hat{\rho} ] + \mathcal{D}[\hat{L}] \hat{\rho},
 \,\,\,\,\,
 \hat L/\sqrt{\kappa} = \hat a_{\0} + \eta \hat a_{\1}^\dagger. \label{eqn:jumpOperator}
\end{align}
Our goal is to understand instabilities and steady states of this setup. Note that previous work studied chiral-symmetric bosonic lattices driven by single-mode squeezing \cite{Yanay2018}. Such systems are completely distinct from our setup: they do not have any dissipative pairing interaction, never exhibit dynamical instability, and (unlike what we describe below) always yield steady states with a {\it spatially uniform} average density.

We start by diagonalizing $\H$. Using chiral symmetry, we can write
$ \H = \sum_{\alpha\geq0} \epsilon_\alpha (\hat d_\alpha^\dagger \hat d_\alpha - \hat d_{-\alpha}^\dagger \hat d_{-\alpha})$. Eigenmode annihilation operators are given in terms of real space wave functions by $\hat d_{\pm \alpha} = \sum_i \psi_{\pm \alpha}[i] \hat a_i$. $\H$ is invariant under two-mode squeezing transformations that mix a pair of $\pm \alpha$ modes \cite{Supplement}: for arbitrary $r_{\alpha}, \phi_\alpha \in \R$, if we take 
\begin{align}
 \hat \beta_{\pm \alpha} &\equiv \cosh(r_\alpha) \hat d_{\pm \alpha} + e^{i \phi_\alpha} \sinh(r_\alpha) \hat d_{\mp \alpha}^\dagger, \label{eqn:betaModes} 
\end{align} 
then $\H = \sum_\alpha \epsilon_\alpha (\hat \beta_\alpha^\dagger \hat \beta_\alpha - \hat \beta_{-\alpha}^\dagger \hat \beta_{-\alpha})$.

We would like to find a set of $r_{\alpha}, \phi_{\alpha}$ such that
\begin{align}
 \hat L &= \sqrt{\kappa} \sum_{\alpha} N_\alpha (\hat \beta_\alpha + \hat \beta_{-\alpha}). \label{eqn:bog_dissipator}
\end{align}
If this is possible, the system dynamics are stable, and we will have a unique steady state (vacuum of the $\hat{\beta}_{\pm \alpha}$ operators). Achieving 
Eq.~(\ref{eqn:bog_dissipator}) requires for each $\alpha>0$ \cite{Supplement}:
\begin{align}
 \tanh r_\alpha &= \eta \left|\frac{\psi_\alpha[\1] }{ \psi_\alpha[\0]^*} \right|, \ \ \ 
 \phi_\alpha = \arg\left(\frac{\psi_\alpha[\1] }{ \psi_\alpha[\0]^*} \right).
 \label{eqn:squeeze_param} 
\end{align}
with $|N_\alpha|^2 = |\psi_\alpha[\0]|^2(1 - |\tanh r_\alpha|^2)$.

We now make a crucial observation: 
\cref{eqn:squeeze_param} only has a solution if $\eta < (|\psi_\alpha[\0]^* / \psi_\alpha[\1]| \equiv \eta_{\alpha})$. If this condition is violated for a particular $\alpha$, then the dynamics is unstable: in this case, we are forced to write $\hat{L}$ in terms of a Bogoliubov raising operator in the $(\alpha, -\alpha)$ sector, implying that the dissipation looks like antidamping in this sector. At a heuristic level, for $\eta > \eta_\alpha$, the $\alpha$ modes see more gain than loss. Overall stability requires $\eta < \textrm{min } \eta_{\alpha} \equiv \eta_c$, a condition that is independent of the dissipation strength $\kappa$. We thus have a generalization and rigorous justification of the surprising FGR-like instability condition in \cref{eqn:instability,eqn:FGR} we found for the three-mode model. 

Our arguments above imply that as long as $\eta < \eta_c$, we are dynamically stable and have a pure steady state, where each $(\alpha,-\alpha)$ pair is in a two-mode squeezed  vacuum with a squeezing parameter given by \cref{eqn:squeeze_param}. This will in general be a highly entangled state. Further, as $\eta \to \eta_c$ from below, the squeezing parameter of the critical modes is diverging, meaning that we will have a pure state where a small subset of modes contribute to a diverging photon number. Note this is very distinct from just incoherent gain and loss, which never has a pure steady state. This behavior is also completely distinct from standard parametric instabilities, where the steady state becomes extremely impure as one approaches instability \cite{Gerry2005,Walls2008}. The mode selectivity leads to a highly nonuniform density that can be exploited for applications, as we now discuss. 

\begin{figure}[t]
 \centering
 \includegraphics[width = 3.25in]{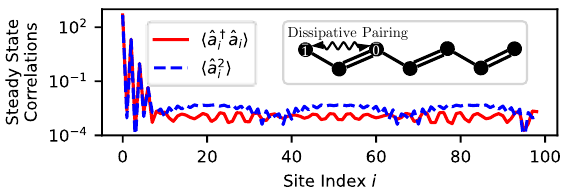}
 \caption{Steady state correlation functions for a 99-site SSH chain with $\delta = -0.65$. There is a single jump operator of the form of \cref{eqn:jumpOperator} with $\0 = 4$ and $\1 = 0$, and $\eta = 0.999\eta_c \sim 0.045$. The squeezing correlation functions show a pure, single-mode squeezed state exponentially localized to the edge. Inset: Schematic of the dissipatively stabilized SSH chain. A single jump operator generates a dissipative pairing interaction, selectively exciting the edge mode.}
 \label{fig:SSH}
\end{figure}
\begin{figure}[t]
 \centering
 \includegraphics[width = 1.4in]{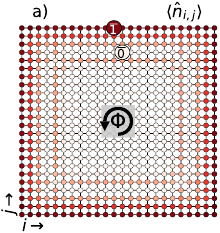}
 \includegraphics[width = 1.4in]{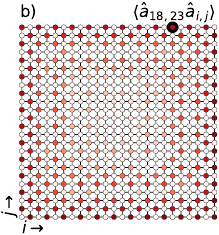}
 \includegraphics[width = 0.38in]{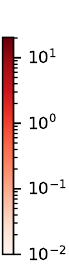}
 \caption{(a) A $24 \times 24$ site Hofstadter lattice, which has uniform hopping and a quarter flux per plaquette $\Phi = \frac{1}{4} \Phi_0$. There is a single dissipator of the form of \cref{eqn:jumpOperator}, with $\1 = (11,23)$ and $\0=(12,20)$, and with $\eta = 0.999 \eta_c \sim 0.0007$. The color corresponds to local steady state photon number, which is exponentially localized to the edges of the lattice. (b) The same system, now showing steady state squeezing correlations between the randomly chosen edge site (18,23) and the rest of the lattice. Every edge site has exponentially enhanced squeezing with every other edge site on the same sublattice.}
 \label{fig:Hofstadter}
\end{figure}

{\textit{Dissipative pairing and topological edge states.---}}
The physics discussed above is particularly striking when applied to chiral hopping Hamiltonians $\H$ that have topological band structures. There are many such models, as chiral symmetry is a key part of the standard classification of topological band structures \cite{Schnyder2008}. As our dissipative interaction always pairs opposite energy modes, edge modes will only be paired with edge modes, bulk modes only with bulk modes. Moreover, it is easy to ensure that the correlated steady-state photon density 
is concentrated on the edges. Edge-mode wave functions are exponentially damped in the bulk, so \cref{eqn:squeeze_param} tells us for an edge state $\alpha$
\begin{align}
 \tanh r_\alpha = \eta \left| \frac{\psi_\alpha[\1]}{\psi_\alpha[\0]} \right| \propto \eta e^{(d_{\0} - d_{\1})/\zeta_L}, \label{eqn:ExpEnhancement}
\end{align}
where $d_{\1,\0}$ is the distance from $\1$ and $\0$ to the edge, respectively, with $\zeta_L$ the localization length scale of the edge modes. If $d_{\0} - d_{\1} > 0$ (i.e.~the gain site closer to the edge than the loss site), we obtain a superexponential enhancement in the squeezing parameter of the edge modes. This yields large populations and squeezing on the edge (while still having a pure state), see \cref{fig:Hofstadter,fig:SSH}.

For large enough systems, the bulk modes will be nearly translationally invariant, implying they will have $\tanh r_\alpha = \eta$. Thus, by spreading the two sites out over a few localization lengths $\zeta_L$, a weak pump rate $\eta \ll 1$ can set $\tanh r_\alpha \sim 1$ for \textit{only} the edge modes. Here, the total number of excitations in the bulk would be very small, $\langle \hat n_\alpha \rangle = O(\eta^2)$, whereas the number of excitations in the edge mode, as one approaches instability, will be superexponentially enhanced and scales like: $\langle \hat n_\alpha \rangle = O([1 - \eta e^{(d_{\0} - d_{\1})/\zeta_L }]^{-1})$.

The upshot is that by using a single dissipative pairing interaction, we can selectively populate, squeeze and entangle edge modes of a topological bosonic band structure. Such states could be useful for applications in topological photonics \cite{Ozawa2019}, and are reminiscent of topological lasing states \cite{Hu2021,Secli2021} (which typically require complex schemes to only pump the edge states).
We analyze this physics more carefully below for two prototypical topological hopping models (see \cref{fig:Hofstadter}). 

{\textit{SSH chain.---}}A paradigmatic topological model is the Su–Schrieffer–Heeger (SSH) chain \cite{Su1979,Su1980}, see \cref{fig:SSH} inset. This is a linear, 1D lattice with staggered hopping strengths, given by the Hamiltonian
\begin{align}
 \H &= -J \sum_{i = 1}^{N - 1} (1 + (-1)^i \delta) \hat a_i^\dagger \hat a_{i + 1} + \mathrm{H.c.}
\end{align}
Such a model has been realized with bosons in a variety of experiments (e.g.~\cite{StJean2017,Parto2018,Zhao2018,Kim2021}). The topological regime of $\H$ admits one (two) protected edge modes if there are an odd (even) number of lattice sites, with a localization length $\zeta_L = (1 + \delta)/(1 - \delta)$. As $\alpha \to -1$, $\zeta_L \to 0$, and the edge modes become infinitely localized. 

We consider for simplicity an odd number of lattice sites (see \cite{Supplement} for even $N$). This yields a single zero-energy edge mode, localized on a single sublattice. Hence, if we place the pairing dissipator on the correct sublattice, we can selectively excite just the edge mode into a single-mode squeezed vacuum with a superexponentially enhanced squeezing parameter. The dissipative steady state for such a situation is plotted in \cref{fig:SSH}. We thus have a resource-friendly approach for creating topologically protected, bright nonclassical squeezed light, using a SSH chain with a single, quasilocal, linear dissipator. One could imagine using the stabilized photons by weakly coupling the edge lattice site to an output waveguide, see \cite{Supplement} for more details.
Note that topological features of the SSH chain are protected against disorder in the hopping coefficients up to the bulk gap $4|\delta|J$. We find that the qualitative nature of the dissipative steady state is also protected against hopping disorder over a similar scale (see \cite{Supplement}). 

{\textit{Hofstadter lattice.---}} 
2D topological systems admit extended boundaries, allowing one to more easily study entanglement properties. Motivated by this, we consider a finite, quarter-flux Hofstadter lattice \cite{Hofstader1976}. This corresponds to a square lattice with a quarter magnetic flux quanta per plaquette (see \cref{fig:Hofstadter}) giving the Hamiltonian:
\begin{align}
 \H &= \sum_{m,n} \hat a_{m,n}^\dagger \hat a_{m + 1,n} + e^{i\pi m/2}\hat a_{m,n}^\dagger a_{m,n + 1} + \mathrm{H.c.},
\end{align}
which has been realized experimentally in Refs. \cite{Hafezi2011,Hafezi2013,Owens2018}.

This Hamiltonian supports exponentially localized modes which propagate chirally around the edge \cite{Hafezi2011,Hafezi2013,Wang2009,TIComment}. The fact that they are extended around the full edge is critical for generating long-range entanglement.

With the same prescription of adding a dissipative interaction of the form of \cref{eqn:jumpOperator} with $\1$ on the edge and $\0$ in the bulk, the steady state solution has exponentially localized edge photon density, with nearly all-to-all edge correlations, \cref{fig:Hofstadter}. For a fixed $\eta$, these sites will obey a volume-law scaling in entanglement entropy, \cite{Supplement}, where maximally separated edge sites are now highly entangled, \cref{fig:Hofstadter}. Having all edge sites lie on the same topological boundary is crucial
for this to occur \cite{Supplement}.

In the limit that $\eta \to \eta_c$, the steady state will be dominated by the topological edge modes approaching instability. Treating the edge as a ring, we can label these by their momenta $k$; the steady state has all momenta $k$ and $k + \pi$ in a TMS vacuum. Close enough to instability, a single momentum will dominate, generating uniform edge photon densities, see \cref{fig:Hofstadter}(a), and a ``checkerboard'' of correlations, see \cref{fig:Hofstadter}(b). The checkerboard is a result of the chiral symmetry, which admits only correlations within a sublattice. The values of the correlations and densities can be understood directly from \cref{eqn:ExpEnhancement}, where $\langle \hat n_{i,j} \rangle \sim \sinh(r_k)^2$ and $\langle \hat a_{i,j} \hat a_{i',j'} \rangle \sim \sinh(r_k) \cosh(r_k)$ are superexponentially enhanced compared to the bulk modes. This gives an arbitrary amount of entanglement between any two edge sites on the same sublattice as $\eta \to \eta_c$. This also means that for a relatively weak dimensionless pumping ($\eta < 10^{-3}$ in \cref{fig:Hofstadter}), the steady state can still have a large number of photons ($O(10^2)$ in \cref{fig:Hofstadter}), that is completely independent of the strength of the dissipation $\kappa$ compared to the Hamiltonian.

{\textit{Implementation.---}} 
The basic master equation is naturally suited for any circuit- or cavity-QED experimental platform that can generate tunable couplings, along with an engineered lossy mode. Quantum systems that have been able to successfully create topological photonic or phononic lattices span superconducting circuits \cite{Owens2018,Kim2021}, micropillar polariton cavities \cite{StJean2017}, photonic cavities \cite{Gong2021}, photonic crystals \cite{Wang2009}, ring resonators \cite{Hafezi2011,Hafezi2013,Leykam2020,Parto2018,Zhao2018}, and optomechanics \cite{Youssefi2022,Peano2015}. In order to generate the requisite jump operator in \cref{eqn:jumpOperator}, one can couple the dissipation sites to an auxiliary bosonic mode $\hat b$ with the interaction
\begin{align}
 \H_I &= g \hat b^\dagger ( \hat a_{\0} + \eta \hat a_{\1}^\dagger) + \mathrm{H.c.}
\end{align}
In the limit that the auxiliary mode $\hat b$ is very lossy with a loss rate $\kappa \gg g$, this gives the desired jump operator, with an effective strength $\Gamma = 4g^2/\kappa$, \cite{Gardiner2004}. This allows one to easily engineer the desired reservoir with few additional resources.

{\textit{Conclusions.---}} 
We have demonstrated that dissipative pairing interactions lead to a previously unexplored class of instabilities in bosonic systems, where stable Hamiltonians and stable dissipation combine to give unstable dynamics. We have shown that these instabilities are incredibly sensitive to topological boundaries, providing a new mechanism to selectively excite topological edge modes without needing any momentum or frequency selectivity. Moreover, the steady state of the dynamics remains pure all the way up to the instability point, allowing one to populate the edge with an arbitrary number of zero-temperature excitations. Our ideas are compatible with a variety of different experimental platforms, and require few resources to implement.

This work is supported by the Air Force Office of Scientific Research under Grant No.~FA9550-19-1-0362, and was partially supported by the University of Chicago Materials Research Science and Engineering Center, which is funded by the National Science Foundation under Grant 
No.~DMR-1420709. A.C. also acknowledges support from the Simons Foundation through a Simons Investigator Award (Grant No.~669487, A.C.).

\nocite{Clerk2010}

\bibliography{ref}

\begin{thebibliography}{62}
\bibitem [{TIC()}]{TIComment}%
  \BibitemOpen
  \href@noop {} {}\bibinfo {note} {The astute reader may be concerned that the
  chiral Altland-Zirnbauer class AIII is topologically trivial in 2D. However,
  the quarter-flux Hofstadter lattice is not gapped at half-filling, instead
  meeting at Dirac cones. The gaps at 1/4 and 3/4 filling have no chiral
  symmetry, and therefore are in class A and topological.}\BibitemShut {Stop}%
\end{thebibliography}

\end{document}


\title{Supplemental Material: Dissipative Pairing Interactions: Quantum Instabilities, Topological Light, and Volume-Law Entanglement}

\author{Andrew Pocklington}
\affiliation{Pritzker School of Molecular Engineering,  University  of  Chicago, 5640  South  Ellis  Avenue,  Chicago,  Illinois  60637,  U.S.A.}

\affiliation{Department of Physics, University  of  Chicago, 5640  South  Ellis  Avenue,  Chicago,  Illinois  60637,  U.S.A. }

\author{Yu-Xin Wang}
\affiliation{Pritzker School of Molecular Engineering,  University  of  Chicago, 5640  South  Ellis  Avenue,  Chicago,  Illinois  60637,  U.S.A.}

\author{A. A. Clerk}
\affiliation{Pritzker School of Molecular Engineering,  University  of  Chicago, 5640  South  Ellis  Avenue,  Chicago,  Illinois  60637,  U.S.A.}

\maketitle

\renewcommand{\theequation}{S\arabic{equation}}
\renewcommand{\thesection}{\Roman{section}}
\renewcommand{\thefigure}{S\arabic{figure}}
\renewcommand{\thetable}{S\arabic{table}}
\renewcommand{\bibnumfmt}[1]{[S#1]}
\renewcommand{\citenumfont}[1]{S#1}



\renewcommand{\theequation}{S\arabic{equation}}
\renewcommand{\thesection}{\Roman{section}}
\renewcommand{\thefigure}{S\arabic{figure}}
\renewcommand{\thetable}{S\arabic{table}}
\renewcommand{\bibnumfmt}[1]{[S#1]}
\renewcommand{\citenumfont}[1]{S#1}

\setcounter{page}{1}
\setcounter{equation}{0}
\setcounter{figure}{0}

\section{Loss-induced instability in 2-site models}
We can investigate unconventional instabilities just at the 2 mode level.
As alluded to in the main text, the dissipative pairing interaction can be thought of as local loss under the proper change of frame. Consider a jump operator:
\begin{align}
    \hat L/\sqrt{\kappa} &= \hat a + \eta \hat b^\dagger
    , 
\end{align}
where we assume $\eta<1$ unless specified otherwise. 
We can make a two-mode squeezing transformation, and define new modes
\begin{align}
    \hat a' &= \frac{1}{\sqrt{1 - \eta^2}} (\hat a + \eta \hat b^\dagger) , \\
    \hat b' &= \frac{1}{\sqrt{1 - \eta^2}} (\hat b + \eta \hat a^\dagger) . 
\end{align}
One can easily check that these obey the canonical commutation relations, and now $\hat L = \sqrt{\kappa(1 - \eta^2)} \hat a'$, giving local loss. However, such a transformation will generically generate active parametric amplifier terms if one starts from a particle-conserving detuning Hamiltonian:
\begin{align}
    \hat a^\dagger \hat a + \hat b^\dagger \hat b &= \frac{1}{1 - \eta^2} (\hat a'^\dagger  - \eta \hat b') (\hat a'  - \eta \hat b'^\dagger)  + \frac{1}{1 - \eta^2} (\hat b'^\dagger - \eta \hat a') (\hat b' - \eta \hat a'^\dagger),  \\
    &= \hat a'^\dagger \hat a' + \hat b'^\dagger \hat b' - \frac{2 \eta}{1 - \eta^2} \left( \hat a' \hat b' + \hat a'^\dagger \hat b'^\dagger \right)
    . 
\end{align}
We believe that the frame presented in the main text is a more intuitive way to understand the dynamics, since all of the steady state features can be read off from the eigenmodes of a passive, hopping Hamiltonian. When we work instead in the frame of the local loss, the importance of the Hamiltonian eigenmodes is no longer as immediate. However, if one insists on working in this frame, we can still explore dynamical instabilities. We can rewrite such a stable parametric amplifier Hamiltonian as:
\begin{align}
    \H &= \Delta(\hat a^\dagger \hat a + \hat b^\dagger \hat b) + \lambda (\hat a^\dagger \hat b^\dagger + \hat a \hat b).
\end{align}
When positive parameters satisfy $\lambda < \Delta$, this Hamiltonian can be diagonalized via a Bogoliubov transformation with eigenmode energies $\pm \sqrt{\Delta^2 - \lambda^2}$, describing stable oscillatory dynamics. Now, let's add some local loss, but only to a single cavity mode. The mode amplitudes thus satisfy linear equations of motion ($a \equiv \langle \hat a \rangle, a^* \equiv \langle a^\dagger \rangle$, etc.):
\begin{align}
    \partial_t \left( \begin{array}{c}
        a \\
        b^* 
    \end{array}
    \right)
    &=
    \left( \begin{array}{cc}
        -i \Delta - \kappa  & -i\lambda \\
        i \lambda & i \Delta
    \end{array}
    \right)
    \left( \begin{array}{c}
        a \\
        b^*  
    \end{array}
    \right). 
\end{align}
Then we find that the eigenvalues of this new problem are $-\frac{\kappa}{2} \pm i \sqrt{\left(\Delta - i \frac{\kappa}{2}\right)^2 - \lambda^2}$. Interestingly enough, one of these always has a positive real part. Moreover, if we consider the more general scenario that we damp both modes with rates $\kappa_a, \kappa_b$, then, defining the sum and difference $\kappa = \frac{\kappa_a + \kappa_b}{2}$ and $\delta \kappa = \frac{\kappa_a - \kappa_b}{2}$, one can show that instability arises whenever:
\begin{align}
    \frac{\delta \kappa^2}{\kappa^2} \geq \frac{\kappa^2 + \Delta^2 - \lambda^2}{\kappa^2 + \Delta^2} . 
\end{align}
Note that this matches our intuition in all the limits we currently have considered: if $\kappa_a = 0$ or $\kappa_b = 0$, then $\kappa = |\delta \kappa|$, and so we \textit{always} have instabilities, since the RHS is always less than 1. If instead we take $\kappa_a = \kappa_b$, then we simply have the standard par-amp instability point at $\lambda^2 = \Delta^2 + \kappa^2$.

This is a seemingly simple instability, that can be seen just at the level of 2 $\times$ 2 matrices. However, it is still surprising, as we have taken a stable Hamiltonian, added to it some local loss, and made it unstable.

To understand more physically the instability mechanism, it is useful to look at the limiting regimes where either the Hamiltonian or the dissipative terms becomes the dominant contribution. 
For concreteness, we focus on the completely unbalanced case, where $\kappa_b = 0$. When $\kappa_a$ is sufficiently small, we can think of it perturbatively as splitting the two eigenmodes. However, the eigenoperators are two-mode squeezed by the Hamiltonian, so for one of the eigenmodes, the loss on mode $a$ acts effectively as cooling, and for the other as heating, giving us instability. When instead $\kappa_a$ becomes dominant, we could adiabatically eliminate $a$ to obtain an effective single-mode model consisting of $b$. In this latter case, it is straightforward to see that $b$ experiences effective heating, since the two original modes are parametrically coupled, again leading to instabilities. 

\begin{figure}[H]
    \centering
    \includegraphics[width = 5.5in]{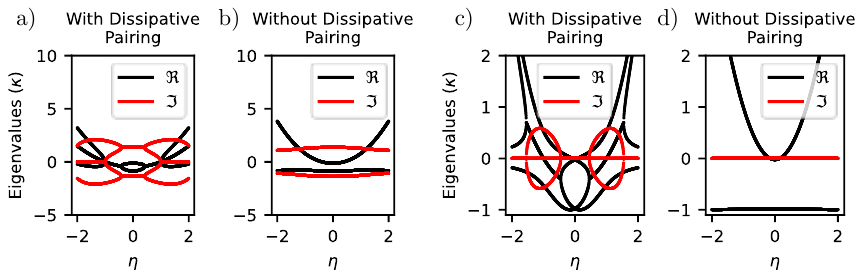}
    \caption{a) and b) show the real and imaginary spectra for the dynamical matrix with a beamsplitter ($J_1 = 0.25 \kappa$) plus parametric amplifier ($J_2 = 0.2 \kappa$) interaction subject to (a) dissipative pairing on the two modes and (b) incoherent gain and loss. Observe that since $\frac{2}{\kappa}\sqrt{J_1^2 - J_2^2} = 0.3$ is less than unity, panel b) depicting uncorrelated dissipation does not have an EP, but the case with correlated dissipation does. c) and d) show the spectra of the dynamical matrix for a detuned beamsplitter Hamiltonian, subject to (c) dissipative pairing on the two modes and (d) incoherent gain and loss. Note that the left exhibits 4 exceptional points (half coming from the $\eta \to - \eta$ symmetry) while the right exhibits none. We have chosen $J = \kappa = \delta = 1$.}
    \label{fig:EPs}
\end{figure}

Now, let's consider a dissipative interaction generated by the jump operator $\hat L/\sqrt{2\kappa} =  \hat a + \eta \hat b^\dagger$. Then adding to this a beam-splitter Hamiltonian $J_1(\hat a^\dagger \hat b + \hat b^\dagger \hat a)$ and a Hermitian parametric amplifier $J_2 (\hat a^\dagger \hat b^\dagger + \hat a \hat b)$, we get a dynamical matrix of the form:
\begin{align}
\partial_t \left( 
\begin{array}{c}
a \\
a^* \\
b \\
b^*
\end{array} 
\right) &= 
\left(
\begin{array}{cccc}
-\kappa & 0 & i J_1 & -i J_2 + \kappa \eta \\
0 & -\kappa & i J_2 + \kappa \eta & -i J_1 \\
i J_1 & -i J_2 - \kappa \eta & \kappa \eta^2 & 0 \\
i J_2 - \kappa \eta & -i J_1 & 0  & \kappa \eta^2
\end{array} 
\right)
\left(
\begin{array}{c}
a \\
a^* \\
b \\
b^* 
\end{array} 
\right) . 
\end{align}
It is interesting to note that if we instead work in the quadratures-like basis $\hat q_{a}^\pm = (e^{i \phi} \hat a \pm \hat a^\dagger)/\sqrt{2}$ and $\hat q_{b}^\pm = (\pm \hat b + e^{i \phi} \hat b^\dagger)/\sqrt{2}$ where $\cos \phi = \frac{J_2}{J_1}$, then we can rewrite the dynamical matrix as 
\begin{align}
\partial_t \left( 
\begin{array}{c}
q_a^+ \\
q_b^+ \\
q_a^- \\
q_b^-
\end{array} 
\right) &= 
\left(
\begin{array}{cccc}
-\kappa  & \sqrt{J_1^2 - J_2^2} + \kappa \eta & 0 & -2i J_2 \\
- \sqrt{J_1^2 - J_2^2} - \kappa \eta & \kappa \eta^2 & 2i J_2 & 0 \\
0 & 0 & -\kappa & -\sqrt{J_1^2 - J_2^2} + \kappa \eta \\
0 & 0 & \sqrt{J_1^2 - J_2^2} - \kappa \eta & \kappa \eta^2
\end{array} 
\right)
\left(
\begin{array}{c}
q_a^+ \\
q_b^+ \\
q_a^- \\
q_b^-
\end{array} 
\right) . 
\end{align}
This dynamical matrix can be equivalently rewritten as
\begin{align}
\left(
\begin{array}{cccc}
-\frac{\kappa}{2} (1 + \eta^2)  & \sqrt{J_1^2 - J_2^2} + \kappa \eta & 0 & -2i J_2 \\
- \sqrt{J_1^2 - J_2^2} - \kappa \eta & \frac{\kappa}{2} (1 + \eta^2) & 2i J_2 & 0 \\
0 & 0 & -\frac{\kappa}{2} (1 + \eta^2) & -\sqrt{J_1^2 - J_2^2} + \kappa \eta \\
0 & 0 & \sqrt{J_1^2 - J_2^2} - \kappa \eta & \frac{\kappa}{2} (1 + \eta^2)
\end{array}
\right) -\frac{\kappa}{2}  (1 - \eta^2) \id ,  \label{seq:DM1}
\end{align}
where we straightforwardly see the $\P \T$-dimer structure in the upper-left and lower-right blocks. Hence, a bit of algebraic manipulation tells us we get an exceptional point (EP) when  
\begin{align}
    \eta &= \pm \left( 1 - \sqrt{\frac{2}{\kappa}} (J_1^2 - J_2^2)^{1/4} \right)
    , 
\end{align}
so that we can always find a value of $\eta$ to realize an EP whenever $J_1^2 > J_2^2$, which occurs at $-\kappa  (1 - \eta^2)/2$. Further, tuning past the EP, we reach an instability point at:
\begin{align}
    \eta &= \frac{\sqrt{\kappa^2 + J_1^2 - J_2^2}-\kappa}{\sqrt{J_1^2 - J_2^2}}
    , 
\end{align}
as can be seen in \cref{fig:EPs}.

We now consider replacing the correlated dissipation with uncorrelated heating and cooling on the two sites. This would be equivalent to setting $ \kappa \eta  = 0$, while leaving $ \kappa \eta ^{2}$ unchanged in \cref{seq:DM1}, which simply shifts the EP to the point where $\frac{\kappa}{2}(1 + \eta^2) = \sqrt{J_1^2 - J_2^2}$. However, this also means that if $\frac{2}{\kappa} \sqrt{J_1^2 - J_2^2} < 1$, the EP disappears under the uncorrelated dissipation, but remains robust for the correlated case. This can be seen in \cref{fig:EPs}.

Another interesting case to consider is instead of a beam-splitter and parametric amplifier, let's consider just a detuned beam-splitter interaction. By setting $J_2 = 0$ in \cref{seq:DM1}, we see that when the detuning is zero, then we would expect an EP. One can show that adding the dissipative interaction makes the EP robust to any detuning, and without it any detuning immediately destroys the EP.

We can similarly write down the dynamical matrix for the case without dissipative interaction, as 
\begin{align}
\partial_t \left( 
\begin{array}{c}
a \\
a^* \\
b \\
b^*
\end{array} 
\right) &= 
\left(
\begin{array}{cccc}
-\kappa  + i \delta & 0 & i J &  \kappa \eta \\
0 & -\kappa - i \delta &  \kappa \eta & -i J \\
i J &  - \kappa \eta & \kappa \eta^2 - i \delta & 0 \\
 - \kappa \eta & -i J & 0  & \kappa \eta^2  + i \delta
\end{array} 
\right)
\left(
\begin{array}{c}
a \\
a^* \\
b \\
b^* 
\end{array} 
\right)
, 
\label{seq:DM2.local}
\end{align}
where $J$ denotes the beam-splitter strength, and $\delta$ detuning between the modes. Interestingly, \cref{seq:DM2.local} actually can have two exceptional points, which occur at
\begin{align}
   \frac{\eta^2}{(\eta^2 - 1)^2} &= \frac{16 J^4 + \kappa^4 + 32 J^2 \delta^2 + 8 \kappa^2 \delta^2 + 16 \delta^4 - 8 J^2 \kappa^2}{64 J^2 \kappa^2} ,  \quad   \text{  and   }
   \quad \frac{\eta^2}{(\eta^2 - 1)^2} = \frac{\delta^2}{4 J^2} .
\end{align}
When $\delta \neq 0$, these exceptional points remain. This can be seen in \cref{fig:EPs}. Without the dissipative interaction, and only incoherent gain and loss, the dynamical matrix takes the block diagonal form:
\begin{align}
\partial_t \left( 
\begin{array}{c}
a \\
b
\end{array} 
\right) &= 
\left(
\begin{array}{cc}
-\frac{\kappa}{2} (1 + \eta^2) + i \delta  & i J  \\
i J & \frac{\kappa}{2} (1 + \eta^2) - i \delta
\end{array} 
\right)
\left(
\begin{array}{c}
a \\
b 
\end{array} 
\right)
- \frac{\kappa}{2}(1 - \eta^2) \id \left(
\begin{array}{c}
a \\
b 
\end{array} 
\right) , 
\end{align}
where we can directly conclude that we create a $\P \T$ dimer and exceptional point if, and only if, we are at zero detuning. Thus, the dissipative interaction generated by the correlated noise makes the EP robust to detuning. This can be seen in \cref{fig:EPs}.

An interesting corollary of our analysis is that two coupled modes subject to unbalanced loss will undergo an EP when the coupling is equal to half the difference of the loss rates. This system cannot, though, ever become unstable.

\section{3 Site Model Dynamical Matrix and Stability Analysis}
Recall our three site Hamiltonian and jump operator:
\begin{align}
    \H &= J_1 \hat a^\dagger \hat b + J_2 \hat b^\dagger \hat c + h.c., \\
    \hat L &= \sqrt{\kappa}( \hat a + \eta \hat c^\dagger).
\end{align}
This Hamiltonian has the unnormalized normal modes
\begin{align}
    \hat d_0 &= J_2 \hat a - J_1 \hat c ,\\
    \hat d_{\pm} &= J_1 \hat a \pm \sqrt{J_1^2 + J_2^2} \hat b + J_2 \hat c , 
\end{align}
which have energies $0, \pm \sqrt{J_1^2 + J_2^2}$. We will call these the dark mode and bright modes, respectively. The time dynamics of the expectation value of an operator $\langle \hat O \rangle$ is given by:
\begin{align}
    \partial_t \langle \hat O \rangle &= i \langle [\H,\hat O] \rangle + \frac{1}{2} \langle \hat L^\dagger [\hat O, \hat L] + [\hat L^\dagger, \hat O] \hat L \rangle
    . 
\end{align}
This gives the semi-classical dynamical matrix for the mode operators expectation values ($a = \langle \hat a \rangle, a^* = \langle a^\dagger \rangle$, etc.):
\begin{align}
\partial_t \left( 
\begin{array}{c}
a \\
b \\
c \\
a^* \\
b^* \\
c^*
\end{array} 
\right) &= 
-i \left(
\begin{array}{cccccc}
-i \frac{\kappa}{2} & J_1 & 0 & 0 & 0 & -i \frac{\kappa}{2} \eta \\
J_1 & 0 & J_2 & 0 & 0 & 0 \\
0 & J_2 & i \frac{\kappa}{2}  \eta^2  & i \frac{\kappa}{2}  \eta & 0 & 0 \\
0 & 0 & -i \frac{\kappa}{2}  \eta & -i \frac{\kappa}{2}   & -J_1 & 0 \\
0 & 0 & 0 & -J_1 & 0 & -J_2 \\
i \frac{\kappa}{2} \eta & 0 & 0 & 0 & -J_2 & i \frac{\kappa}{2}  \eta^2
\end{array} 
\right)
\left(
\begin{array}{c}
a \\
b \\
c \\
a^* \\
b^* \\
c^*
\end{array} 
\right) . 
\end{align}
Before we calculate the stability conditions for this matrix, we can gain some intuition by performing a simple perturbative calculation. If we know the eigenvectors and eigenvalues of a matrix $M$ satisfy $M|n\rangle = \epsilon_n |n\rangle$, and we are interested in the spectrum of $M + \delta M$, then to first order, this is simply $\epsilon_n + \langle n | \delta M | n \rangle$. For stability, we only care about the real part, and so we can simply calculate $ \frac{1}{2} \langle n | \delta M + \delta M^\dagger | n \rangle$. For our system, the dissipative interaction is \textit{Hermitian}, so to leading order it doesn't affect the stability conditions at all. Hence, we can just focus on the local heating and cooling rates in the perturbative analysis, which tells us that for the bright modes, we expect a relaxation rate proportional to $\kappa(J_1^2 - \eta^2 J_2^2)$, and for the dark mode a relaxation rate proportional to $\kappa (J_2^2 - \eta^2 J_1^2)$. If either is negative, the system would be dynamically unstable, giving rise to exponential growth, which occurs exactly at the point $|\eta| > |J_1/J_2|$ for the bright modes, and $|\eta| > |J_2/J_1|$ for the dark mode, reproducing the values quoted in the main text. 

To show that these are actually the instability points for all $\kappa$, we must go beyond this leading order calculation. Fortunately, we can observe that, because the dissipation is confined to a single sublattice, flipping the sign of the other sublattice - that is $\hat b \to -\hat b$, is equivalent to sending $J_i \to - J_i$ (in this sense, the dissipator respects the chiral sublattice symmetry of the Hamiltonian). Next, because the Hamiltonian terms are all imaginary, and the dissipative terms all real, complex conjugation $a \to a^*$, etc, also sends $J_i \to -J_i$. In other words, the map $a \to a^*, b \to -b^*, c \to c^*$ constitutes a symmetry of the full dynamical matrix. This allows us to block diagonalize in terms of the quadratures $x_a = \langle \hat a  + \hat a^\dagger \rangle /\sqrt{2}, p_b = i\langle \hat b^\dagger  - b \rangle /\sqrt{2}$ and $x_c = \langle \hat c + \hat c^\dagger \rangle /\sqrt{2}$:
\begin{align}
\partial_t \left( 
\begin{array}{c}
x_a \\
p_b \\
x_c 
\end{array}
\right)
&=  -i \left(
\begin{array}{ccc}
-i \frac{\kappa}{2} &  i J_1  & - i \frac{\kappa}{2} \eta \\
- i J_1 & 0 & - i J_2  \\
 i \frac{\kappa}{2} \eta & i J_2 & i \frac{\kappa}{2} \eta^2
\end{array} 
\right)
\left( 
\begin{array}{c}
x_a \\
p_b \\
x_c 
\end{array}
\right) . 
\label{seq:DM.3m}
\end{align}

We can analytically calculate the spectrum of \cref{seq:DM.3m}, as the roots of the characteristic equation of the dynamical matrix (setting $\kappa = 2$ for simplicity):
\begin{align}
x^3 + (1 - \eta^2)x^2 + J^2 x + \Delta = 0 .  \label{seqn:roots}
\end{align}
where $J^2 = J_1^2 + J^2_2$ and $\Delta = J_2^2 - \eta^2 J_1^2$. We can break this into an equation of motion for the real and imaginary parts by defining $x = v + iw$ to get
\begin{align}
\label{seq:char.3m.re}
v^3 - 3 vw^2 + (1 - \eta^2)(v^2 - w^2) + J^2 v + \Delta &= 0 ,\\
\label{seq:char.3m.im}
-w^3 + 3 v^2 w + 2(1 - \eta^2) vw + J^2 w &= 0 . 
\end{align}
There are two solutions for the imaginary part, $w = 0$ or $w^2 = 3v^2 + 2(1 - \eta^2)v + J^2$. We can identify the first $w = 0$ term with the dark mode, as it has zero energy, and it is only dissipative. Plugging this back in \cref{seq:char.3m.re}, we find the real part then satisfies
\begin{align}
v^3 + (1 - \eta^2)v^2 + J^2 v + \Delta = 0 , 
\end{align}
which has a positive root if, and only if,
\begin{align}
    \Delta < 0 \iff |\eta| > \left|\frac{J_2}{J_1} \right|,
\end{align}
which is exactly the stability condition predicted from a FGR calculation, reproducing the result in the main text.

We can get a very similar picture for the other branch of solution to \cref{seq:char.3m.im}. If we take the imaginary part to be $w^2 = 3v^2 + 2v + J^2$, which is exactly the energy of the bright modes, plus a shift coming from the dissipation, then we get an equation for the relaxation rates:
\begin{align}
-8 v^3 - 8(1 - \eta^2) v^2 - 2( J^2 + (\eta^2 - 1)^2)v - J^2(1 - \eta^2) + \Delta &= 0
. 
\end{align}
This has a positive real root if and only if the following condition holds: 
\begin{align}
    \Delta > J^2(1 - \eta^2) \iff |\eta| > \left|\frac{J_1}{J_2} \right|.
\end{align}
This is again exactly the FGR value for the bright modes.

Finally, we consider the case when we lose the dissipative interaction. Then, creation and annihilation operators decouple, so we need only solve the spectrum of:
\begin{align}
D &=  \left(
\begin{array}{ccc}
-i \frac{\kappa}{2}  & -J_1  & 0 \\
-J_1 & 0 & -J_2  \\
0 & -J_2 & i \frac{\kappa}{2} \eta^2 
\end{array} 
\right),
\end{align}
where we have reintroduced the rate $\kappa$. Now, the eigenvalues correspond to solutions of the cubic equation:
\begin{align}
    x^3 + \frac{\kappa}{2} (1 - \eta^2) x^2 + \left( J^2 - \frac{\kappa^2 \eta^2}{4}  \right)x + \frac{\kappa}{2} \Delta = 0 . 
\end{align}
In the limit of weak dissipation, $\kappa \eta \ll J$, the roots are identical to those of \cref{seqn:roots}. (For the dark mode, they are equivalent all the way up to $|\kappa \eta| = 2|J|$, afterwards there is a sudden shift.) However, as soon as this is not the case, the instability point can move in a $\kappa$ dependent manner. The interaction is perfectly balanced in \cref{seq:DM.3m} such that the instability points in that case are completely independent of $\kappa$.

\section{Quantum Master Equation Steady State}
If we start from the full lattice Hamiltonian and jump operator
\begin{align}
    \H &= \sum_{ij} H_{ij} \hat a_i^\dagger \hat a_j + h.c., \\
    \hat L/\sqrt{\kappa} &= \hat a_{\0} + \eta \hat a_{\1}^\dagger,
\end{align}
then we can diagonalize the Hamiltonian into normal modes $\hat d_{\pm \alpha} = \sum_i \psi_{\pm \alpha}[i] \hat a_i$. Here, we have defined the unitary transformation $\psi_\alpha[i]$ between real space lattice site operators $\hat a_i$ and eigenmode operators $\hat d_\alpha$. Taking advantage of the fact that there is a real space chiral sublattice symmetry, we have  
\begin{align}
    \H &= \sum_{\alpha > 0} \epsilon_\alpha (\hat d_\alpha^\dagger \hat d_\alpha - \hat d_{-\alpha}^\dagger \hat d_{-\alpha}) , \\
    \psi_{-\alpha}[i] &= (-1)^i \psi_\alpha[i] . 
\end{align}
Next, observe that such a chiral Hamiltonian is invariant under two-mode squeezing transformations between $\hat d_\alpha$ and $\hat d_{-\alpha}$, constituting a massive $\SU(1,1)^{N/2}$ symmetry group. That is, we can define the new operators $\hat \beta_{\pm \alpha} = \cosh(r_\alpha) \hat d_{\pm \alpha} + e^{i \phi_\alpha} \sinh(r_\alpha) \hat d_{\mp \alpha}^\dagger$ with energy-dependent squeezing parameters $r_\alpha$. The Hamiltonian in this basis becomes:
\begin{align}
    \H &= \sum_{\alpha > 0} \epsilon_\alpha (\hat \beta_\alpha^\dagger \hat \beta_\alpha - \hat \beta_{-\alpha}^\dagger \hat \beta_{-\alpha}), 
\end{align}
which is still diagonal. Further, expanding our jump operator in eigenmodes, we find
\begin{align}
    \hat L/\sqrt{\kappa}  &=  \hat a_{\0} + \eta \hat a_{\1}^\dagger \\
    &= \sum_{\alpha > 0} \left [ \left( \psi_\alpha[\0]^* \hat d_\alpha + (-1)^{\0} \psi_{\alpha}[\0]^* \hat d_{-\alpha} \right) + \eta \left( \psi_\alpha[\1] \hat d_\alpha^\dagger + (-1)^{\1} \psi_{\alpha}[\1] \hat d_{-\alpha}^\dagger \right) \right] . 
\end{align}
Now, we assume $\0,\1$ live on the same sublattice, and, without loss of generality, we define $(-1)^{\0} = (-1)^{\1} = 1$. Thus, we can express $\hat L$ as
\begin{align}
    \hat L /  \sqrt{\kappa} 
    &= \sum_{\alpha > 0} \left[\left(  \psi_\alpha[\0]^* \hat d_\alpha + \eta \psi_{\alpha}[\1] \hat d_{-\alpha}^\dagger   \right) +  \left( \psi_{\alpha}[\0]^* \hat  d_{-\alpha} + \eta \psi_\alpha[\1] \hat d_\alpha^\dagger \right) \right]  \label{seqn:dissipator} \\
    &\equiv \sum_{\alpha > 0} N_\alpha (\hat \beta_\alpha + \hat \beta_{-\alpha}),
\end{align}
where we define the squeezing parameters and normalizations:
\begin{align}
    \tanh r_\alpha &= \eta  \left| \frac{\psi_\alpha[\1]}{\psi_\alpha[\0]^*} \right| ,\\ \phi_\alpha &= \arg \left( \frac{\psi_\alpha[\1]}{\psi_\alpha[\0]^*}  \right),\\
    |N_\alpha|^2 &= |\psi_\alpha[\0]|^2  - \eta^2 |\psi_\alpha[\1]|^2  .
\end{align}
Thus, the placement of the dissipation sites sets the squeezing and a phase reference for the lattice.

One can now straightforwardly see that, since the Hamiltonian is diagonal in the Bogoliubov modes $\hat \beta$, and the jump operator is purely cooling, the steady state is uniquely the joint vacuum of all the Bogoliubov modes.

As noted in the main text, if $\tanh r_\alpha > 1$, we no longer have a well defined Bogoliubov transformation. This corresponds to the case that $|\psi_\alpha[0]| < |\eta \psi_\alpha[\1]|$. In such a case, to define a squeezing transformation that respects the canonical commutation relations, then we must define 
\begin{align}
    \hat \beta_{\pm \alpha} \propto \eta \psi_\alpha[\1]^* \hat d_{-\alpha} + \psi_\alpha[\0] \hat d_\alpha^\dagger.
\end{align}
If we define the sets $A = \{ \alpha > 0 \ | \ |\psi_\alpha[\0]| > |\eta \psi_\alpha[\1]| \} $ and $B = \{ \alpha > 0 \ | \  |\psi_\alpha[\0]| < |\eta \psi_\alpha[\1]| \} $, then \cref{seqn:dissipator} tells us 
\begin{align}
    \hat L/\sqrt{\kappa} &= \sum_{\alpha \in A} N_\alpha (\hat \beta_\alpha + \hat \beta_{-\alpha}) + \sum_{\alpha \in B} N_\alpha (\hat \beta_\alpha^\dagger + \hat \beta_{-\alpha}^\dagger).
\end{align}
Hence, as soon as $B$ is non-empty, at least one eigenmode is experiencing heating, which means the dynamics are unstable.

We can finally observe that these instability points occur at
\begin{align}
    |\eta| > \min_{\alpha} 
    \left| \frac{\psi_\alpha[\0]}{\psi_\alpha[\1]} \right|,
\end{align}
which are exactly the FGR-based predictions expected from first-order perturbation theory, just as in the three-site model.

\section{Steady State with Squeezed Vacuum Noise}

In this section, we will consider the steady state if, instead of the dissipative pairing interaction, we simply couple the lattice to squeezed vacuum noise. Specifically, we will demonstrate that it is impossible to generate instability, and also that there is no useful spatial selectivity. 

To see that it is impossible to generate instability, we can begin by just looking at the dynamical matrix. If we have two dissipators $\hat L_1 = \sqrt{\kappa}(\hat a_{\0} + \eta \hat a_{\1}^\dagger)$ and $\hat L_2 = \sqrt{\kappa}(\hat a_{\1} + \eta \hat a_{\0}^\dagger)$, then 
\begin{align}
    \partial_t \langle \hat a_{\0} \rangle &= \frac{1}{2}\sum_{i = 1,2} \langle \hat L_i^\dagger [\hat a_{\0}, \hat L_i] + [\hat L_i^\dagger, \hat a_{\0}] \hat L_i \rangle \\
    &= -\frac{\sqrt{\kappa}}{2} \left\langle \hat L_1 - \eta \hat L_2^\dagger \right\rangle = -\frac{\kappa}{2} \left( 1 - \eta^2 \right) \langle \hat a_{\0} \rangle . 
\end{align}
Similarly, the dynamics is symmetric under $\hat a_{\0} \leftrightarrow \hat a_{\1}$ if we correspondingly exchange $\hat L_1 \leftrightarrow \hat L_2$, so we have 
\begin{align}
    \partial_t \langle \hat a_{\1} \rangle &= -\frac{\kappa}{2} \left( 1 - \eta^2 \right) \langle \hat a_{\1} \rangle. 
\end{align}
Taking a generic Hamiltonian $\H = \sum_{i,j} H_{i,j} \hat a_i^\dagger \hat a_j$, we get the equations of motion 
\begin{align}
    \partial_t \langle \hat a_i \rangle &= \sum_j \left(-i H_{i,j}  - \frac{\kappa}{2}(1 - \eta^2)(\delta_{j,\0} + \delta_{j,\1}) \delta_{i,j} \right) \langle \hat a_j \rangle \\
    &\equiv -i\sum_j D_{i,j}  \langle \hat a_j \rangle
    , 
\end{align}
where we have defined the dynamical matrix $D_{i,j} = D_{i,j}^H + D_{i,j}^\kappa$, with
\begin{align}
    D_{i,j}^H &= H_{i,j} , \\
    D_{i,j}^\kappa &= -i\frac{\kappa}{2}(1 - \eta^2)(\delta_{j,\0} + \delta_{j,\1}) \delta_{i,j} . 
\end{align}
Now, if the system is unstable, then there must be an eigenvalue $\lambda$ of $D$ such that $\mathrm{Im} \lambda > 0$. Let's associate such an eigenvalue with an eigenvector $|\psi \rangle$, implying
\begin{align}
    \mathrm{Im} \lambda &= -\frac{i}{2} \langle \psi | D - D^\dagger | \psi \rangle = -i \langle \psi | D^\kappa | \psi \rangle.
\end{align}
However, $-iD^\kappa$ is negative semi-definite, a contradiction that $\mathrm{Im} \lambda$ is positive.

Next, we will prove that, if there is a pure steady state, there is uniform photon density in the lattice. To see this, note that $\hat L_1$ and $\hat L_2$ are both in the same form as the dissipator we have so far considered, so for an arbitrary Hamiltonian, we know that we can write them as a sum of Bogoliubov modes. Hence, we know that 
\begin{align}
    \hat L_1/\sqrt{\kappa} &= \sum_\alpha N_\alpha \left(\hat \beta_\alpha + \hat \beta_{-\alpha} \right) , 
\end{align}
with
\begin{align}
    \tanh r_\alpha &= \left| \eta \frac{\psi_\alpha[\1]}{\psi_{\alpha}[\0]^*} \right|. \label{seq:tanh1}
\end{align}
We can write $\hat L_2$ in the same form, up to exchanging $\0 \leftrightarrow \1$, i.e.
\begin{align}
    \tanh r_\alpha &= \left| \eta \frac{\psi_\alpha[\0]}{\psi_{\alpha}[\1]^*} \right|. \label{seq:tanh2}
\end{align}
We know that the steady state is unique and pure with just $\hat L_1$, hence, for it to remain pure after adding $\hat L_2$, the dark state must also be annihilated by $\hat L_2$. Equating \cref{seq:tanh1} and \cref{seq:tanh2} gives
\begin{align}
    \implies |\psi_\alpha[\1]|^2 &= |\psi_{\alpha}[\0]|^2, \\
    \implies |\tanh r_\alpha| &= \eta, \\
    \implies \langle a_i^\dagger a_i \rangle &= \frac{\eta^2}{1 - \eta^2} \ \forall i. 
\end{align}
Thus, to get a pure steady state, it is necessary and sufficient that there exists some symmetry of the Hamiltonian that guarantees $|\psi_\alpha[\1]|^2 = |\psi_{\alpha}[\0]|^2$ for every eigenmode. In this case, the squeezing parameter for every pair of eigenmodes is identical to $\eta$, meaning one cannot get any topological selectivity, and in fact there is always a uniform photon density in the lattice.

If such a symmetry does not exist, the steady state is guaranteed to be in an impure state.

\section{Topological Enhancement}

\subsection{SSH Degeneracy}
In the main text, we considered an SSH chain with an \textit{odd} number of lattice sites, because an even-numbered SSH chain will have two edge modes whose energies go to zero exponentially quickly in system size. When there are energy degeneracies, the steady state is no longer unique; this means that the time it would take to prepare the steady state is growing exponentially in system size, severely limiting the practical use.

First, we prove explicitly that the dissipative gap closes when the Hamiltonian is degenerate. To begin, we will assume that we have a generic Hamiltonian that satisfies the proper symmety conditions. That is, it can be unitarily diagonalized into modes $\hat d_{\pm \alpha} = \sum_i \psi_\alpha[i] \hat a_i$ where $\psi_{-\alpha}[i] = (-1)^i \psi_\alpha[i]$ and the energy of $\hat d_\alpha$ is opposite that of $\hat d_{-\alpha}$. 

As before, we assume we have a single jump operator of the form $\hat L/\sqrt{\kappa} = \hat a_{\0} + \eta \hat a_{\1}^\dagger$. From here, we can do the transformations:
\begin{align}
	\H &= \sum_{\alpha > 0} \epsilon_\alpha (\hat d_\alpha^\dagger \hat d_\alpha - \hat d_{-\alpha}^\dagger \hat d_{-\alpha}) \\
	& \equiv \sum_{\alpha > 0} \epsilon_\alpha (\hat \beta_\alpha^\dagger \hat \beta_\alpha - \hat \beta_{-\alpha}^\dagger \hat \beta_{-\alpha}) ,\\
	\hat L &= \sqrt{\kappa} 
	(\hat a_{\0} + \eta \hat a_{\1}^\dagger) \\
    &= \sqrt{\kappa} \sum_{\alpha > 0} \left[ \left( \psi_\alpha[\0]^* \hat d_\alpha + (-1)^{\0} \psi_{\alpha}[\0]^* \hat d_{-	\alpha} \right) + \eta \left( \psi_\alpha[\1] \hat d_\alpha^\dagger + (-1)^{\1} \psi_{\alpha}[\1] \hat d_{-\alpha}^\dagger \right) \right ] \\
    &= \sqrt{\kappa} \sum_{\alpha > 0} \left [\left(  \psi_\alpha[\0]^* \hat d_\alpha + (-1)^{\0} \eta \psi_{\alpha}[\1] \hat d_{-\alpha}^\dagger   \right) +  (-1)^{\0} \left( \psi_{\alpha}[\0]^* \hat  d_{-\alpha} + (-1)^{\0} \eta \psi_\alpha[\1] \hat d_\alpha^\dagger \right) \right ] \\
    &\equiv \sqrt{\kappa} \sum_{\alpha > 0} N_\alpha (\hat \beta_\alpha + (-1)^{\0}\hat \beta_{-\alpha}),
\end{align}
where we used that $\0$ and $\1$ are on the same sublattice, so $(-1)^{\0} = (-1)^{\1}$. The system thus has a unique steady state \textit{as long as the spectrum is non-degenerate}. If the spectrum has two degenerate modes -- lets suppose $\epsilon_{\gamma} = \epsilon_{\delta}$ -- then we can always rotate between them to define the new (unnormalized) modes
\begin{align}
    \hat \beta_{\gamma}' &= N_{\gamma} \hat \beta_{\gamma} + (-1)^{\0} N_{\delta} \hat \beta_\delta,  \\  
    \hat \beta_{\delta}' &= N_{\delta} \hat \beta_{\gamma} - (-1)^{\0} N_{\gamma} \hat \beta_\delta .  \label{seqn:slack_mode}
\end{align}
These are equally good Hamiltonian eigenmodes, but the dissipation only cools $\hat \beta_{\gamma}'$ and not $\hat \beta_{\delta}'$. Thus, the steady state cannot be unique. 

In the SSH chain, the topological edge modes become degenerate zero modes exponentially quickly as $N \to \infty$, and therefore the time it takes to reach the steady state is blowing up exponentially in the system size, making one exponentially susceptible to e.g. photon loss. There is a simple solution to this problem, whereby one can always cool the slack mode [defined by \cref{seqn:slack_mode}] in any lattice with mirror symmetry by adding the mirrored dissipator. Here, ``mirror symmetry'' means the inversion symmetry about the center of the lattice, sending $\hat a_i \leftrightarrow \hat a_{N - 1 - i}$ (if the lattice is zero-indexed).

This symmetry means that adding a second dissipator $\hat L'/\sqrt{\kappa} = \hat a_{N - 1 - \0} - \eta \hat a_{N - 1 - \1}^\dagger$ will give the exact same effective squeezing parameters $r_\alpha$, so it has the same dark state. However, it lives on the opposite sublattice as $\hat L$, and so there is a phase flip - this can be seen most clearly by simply expanding in the proper mode basis:
\begin{align}
\hat L' &= \sqrt{\kappa}(\hat a_{N - 1 - \0} - \eta \hat a_{N - 1 - \1}^\dagger) \label{seqn:mirrored_diss} \\
	&= \sqrt{\kappa}\sum_{\alpha > 0} \left[ 
	\psi_\alpha[N - 1 - \0]^*
	\left(  \hat d_\alpha + (-1)^{N - 1 - \0}  \hat d_{-	\alpha} \right) - \eta
	\psi_\alpha[N - 1 - \1]
	\left(  \hat d_\alpha^\dagger + (-1)^{N - 1 - \1}  \hat d_{-\alpha}^\dagger \right) \right ].
\end{align}
Here, it is useful to observe that $N$ is even, so $(-1)^{N - 1 - \0} = (-1)^{1 - \0}$. Further, the mirror symmetry tells us explicitly that $\psi_\alpha[i] = \psi_\alpha[N - 1 - i]$, allowing us to make the simplificattion:
\begin{align}
   \hat L' 	&= \sqrt{\kappa}\sum_{\alpha > 0} \left [ \left( \psi_\alpha[\0]^* \hat d_\alpha + (-1)^{1 - \0} \psi_{\alpha}[\0]^* \hat d_{-	\alpha} \right) - \eta \left( \psi_\alpha[\1] \hat d_\alpha^\dagger + (-1)^{1 - \0} \psi_{\alpha}[\1] \hat d_{-\alpha}^\dagger \right) \right] \\
   &= \sqrt{\kappa} \sum_{\alpha > 0} \left [
   \left(  \psi_\alpha[\0]^* \hat d_\alpha - (-1)^{1 - \0}\eta \psi_{\alpha}[\1] \hat d_{-\alpha}^\dagger   \right) +  (-1)^{ 1 - \0} \left( \psi_{\alpha}[ \0]^* \hat  d_{-\alpha} - (-1)^{1 - \0}\eta \psi_\alpha[\1] \hat d_\alpha^\dagger \right) 
   \right] \\
    & \equiv \sqrt{\kappa} \sum_{\alpha > 0} N_\alpha (\hat \beta_\alpha + (-1)^{1 - \0}\hat \beta_{-\alpha}),
\end{align}
Since $(-1)^{ 1 - \0}(-1)^{\0} = -1$ is negative, between $\hat L$ and $\hat L'$ they cool both the sum and difference of the two degenerate zero modes, relieving the problem of the degeneracy. 

\begin{figure}[H]
    \centering
    \includegraphics[width = 3in]{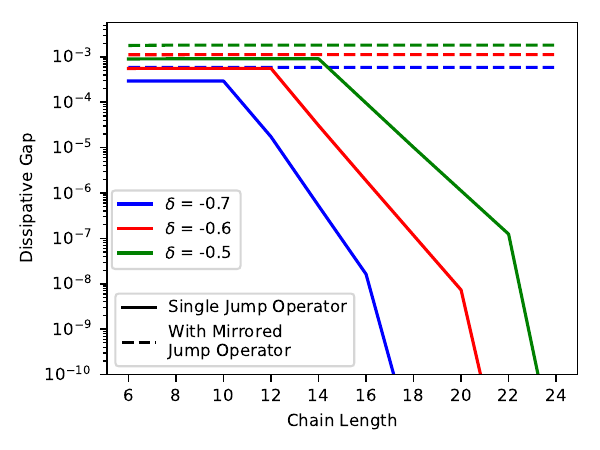}
    \caption{This shows the dissipative gap, or the relaxation rate of the slowest mode, for an SSH chain at $\eta = 0.99 \eta_c$, $\kappa = J = 1$, and $\0 = 2$, $\1 = 0$. Note that when we have a single jump operator, the dissipative gap decays exponentially in system size, meaning the stabilization time scale is quickly diverging. On the other hand, adding a second mirrored dissipator in the form of \cref{seqn:mirrored_diss} removes this exponential suppression.}
    \label{fig:SSH_Gap}
\end{figure}

To see this in action, one can observe that even for relatively small lattices, adding the second dissipator can increase the dissipative gap by several orders of magnitude, see \cref{fig:SSH_Gap}.

\begin{figure}[H]
    \centering
    \includegraphics{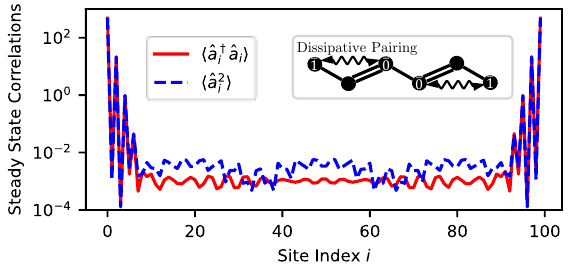}
    \caption{This shows the steady state correlations for an even-numbered (100 site) SSH chain. We have taken $\delta = -0.7$, and two dissipators: one has $\0 = 4$ and $\1 = 0$, the mirrored one takes $\0 = 95$, $\1 = 99$. Both have $\eta = 0.999\eta_c$. The steady state solution shows single mode squeezing of each edge.}
    \label{fig:SSH_even}
\end{figure}

For such an even-numbered SSH chain, the steady state will still be single-mode squeezed vacuums of the edges. To see this, we can observe that the edge eigenmodes that obey the real-space chiral symmetry condition are not localized, but are superpositions of localized edge modes. If we let $|\psi_L \rangle$ be left-localized and $|\psi_R\rangle$ be right localized, then the edge eigenmodes are $|\psi_{\pm} \rangle = (|\psi_L \rangle \pm |\psi_R \rangle )/\sqrt{2}$. The steady state will therefore be a two mode squeezed vacuum of $|\psi_+ \rangle$ and $|\psi_-\rangle$. However, two mode squeezing plus such a beam splitter operation is equivalent to simple having single-mode squeezing on both modes. This can be seen in \cref{fig:SSH_even}.

\subsection{SSH Topological Protection}

Here, we show that the SSH chain has topological protection. To do this, we will add disorder to coupling strengths of the SSH chain, whose spectrum should be robust to perturbations whose strength is less than the bulk gap $4|\delta|J$. For each steady state, we will define a symmetrized edge localization factor:
\begin{align}
    \mathcal{S} &= \frac{1}{n}\sum_i \sqrt{n_i n_{N - 1 - i}} \left( 1 -  4\frac{i (N - i)}{N^2} \right), \\
    n_i &= \langle \hat a_i^\dagger \hat a_i \rangle ,\\
    n &= \sum_i n_i . 
\end{align}
It has been defined in this manner so that the edge localization value goes to unity if the particles are infinitely localized symmetrically to the edges, and goes to zero if they are localized to the center of the bulk or localized to a single edge. The reason for this is that as the disorder gets large, the wavefunctions undergo Anderson localization, so if we have pumping on the edge lattice site, the steady state photon number tends to be concentrated there, but the other edge mode is not excited. On the other hand, when there is topological protection and two edge modes, there should be photon density evenly distributed to both sides of the lattice.

\begin{figure}[H]
    \centering
    \includegraphics[width = 2.5in]{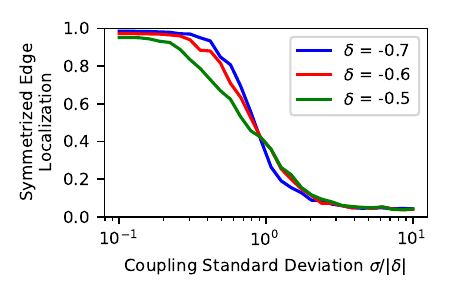}
    \includegraphics[width = 2.5in]{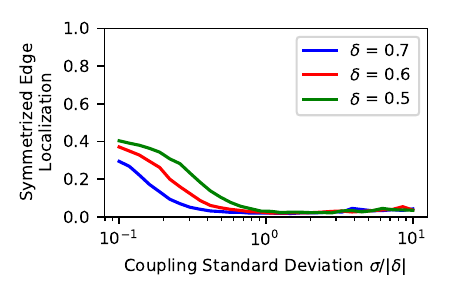}
    \caption{Plotted on the left (right) is an $N = 20$ site SSH chain, with $\delta < 0$ ($\delta > 0$) in a topologically nontrivial (trivial) regime. In the topological regime, for all three values of $|\delta|$ shown, there is a clear crossover at $\sigma \sim |\delta|$ from topologically protected edge excitations with $\mathcal{S} \sim 1$, to Anderson localization with $\mathcal{S} \sim 0$. The topologically trivial regime has no edge modes, transitioning from extended bulk excitations when $\sigma \sim 0$ to Anderson localization when $\sigma \gg \delta$. Both plots have a single dissipator with $\kappa = 1$, $\0 = 2$, $\1 = 0$, and $\eta = 0.99 \eta_c$. Data is averaged over 500 disorder realizations.}
    \label{fig:SSH_Protection}
\end{figure}

In \cref{fig:SSH_Protection}, we can observe that for a range of $\delta$ values, the steady state value of the symmetric edge localization is protected up to disorder $\sigma \sim |\delta|$, where the SSH Hamiltonian is
\begin{align}
    \H &= \sum_{i = 0}^{N - 2} [1 - (-1)^{i} \delta + J_i] \hat a_i^\dagger \hat a_{i + 1} + h.c. , 
\end{align}
and the $J_i$ are sampled from a Gaussian distribution with zero average and width $\sigma$. 

At this point, we would like to stress that the topology only protects the steady state to Hamiltonian perturbations which conserve the chiral symmetry. If one were instead interested in the effect of other perturbations, e.g. a finite lifetime of the local bosonic modes, then one would have to consider the full Liouvillian spectrum, where the Liouvillian $\L$ is the superoperator which governs the dynamics of the full density matrix, i.e.
\begin{align}
    \partial_t \hat \rho &= \L \rho = -i[\H,\hat \rho] + \D[\hat L] \hat \rho .
\end{align}
In this case, the Liouvillian gap is closing quickly as $N$ gets large, which can be understood intuitively from the fact that we are using a single jump operator to stabilize the entire lattice, and so it takes time for excitations on top of the Bogoliubov vacuum to propagate to the dissipation sites and then decay away. 

Hence, there is no protection to additional noise. In some sense, we have chosen the most resource-efficient scheme possible by using a single, quasi-localized jump operator. The price paid by this choice is that the single jump operator must be stronger than all of the other possible sources of relaxation in the lattice. If one considers preparing the steady state quickly more important than resource efficiency, one can imagine a continuum of possible models whereby one adds a number of other, compatible dissipators throughout the lattice, trading resources for speed.

\subsection{SSH Output Fields}

\begin{figure}[H]
    \centering
    \includegraphics[width = 6in]{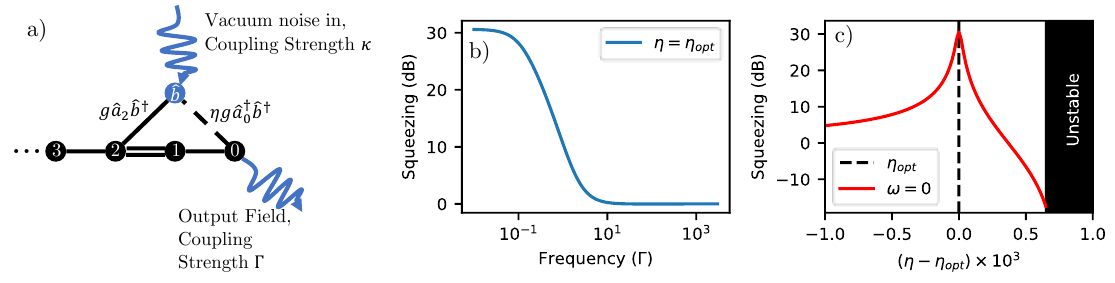}
    \caption{a) depicts a schematic of an SSH chain, along with an auxilliary lattice site $\hat b$ to generate the desired dissipation. Input vacuum noise on $\hat b$ drives the system to the steady state, which is then allowed to leak out of the system from site $0$ on the edge. b) and c) show the squeezing spectrum with $\Gamma = 10^{-3}, \kappa = 10, g = 4, J = 1, \delta = -0.65$, and $\eta_{opt} \sim 0.212$ vs (b) frequency and (c) $\eta$. $\eta_{opt}$ is defined as the optimal imbalance $\eta$ to maximize output squeezing at zero frequency. This is slightly less than $\eta_c$, where the system becomes unstable, with instability denoted by black shading in (c). }
    \label{fig:SSH_Output}
\end{figure}

In this section, we analyze the output fields generated by weakly tapping off some light from the topological edge mode. We can imagine generating the dissipative interaction by coupling to a heavily damped ancilla mode $\hat b$, 
\begin{align}
    \H &= \sum_{i,j} H_{i,j} \hat a_i^\dagger \hat a_j + g(\hat a_{\0} + \eta \hat a_{\1}^\dagger) \hat b^\dagger + g(\hat a_{\0}^\dagger + \eta \hat a_{\1}) \hat b . 
\end{align}
If the $\hat b$ mode is lossy with a decay rate $\kappa$, and we also couple the edge lattice site $\hat a_0$ to a waveguide with a rate $\Gamma \ll \kappa$, see \cref{fig:SSH_Output}(a), the Heisenberg-Langevin equations of motion will be:
\begin{align}
    \partial_t \hat a_i &= -i[\H,\hat a_i] + \delta_{i,0} \left(-\frac{\Gamma}{2} \hat a_0 + \sqrt{\Gamma} \hat a_{in} \right),  \\
    \partial_t \hat b &= -i[\H,\hat b] -\frac{\kappa}{2} \hat b + \sqrt{\kappa} \hat b_{in} , 
\end{align}
where $\hat a_{in}, \hat b_{in}$ are operator valued, zero-temperature white noise: $\langle \hat a_{in}(t) \hat a_{in}(t')^\dagger \rangle = \langle \hat b_{in}(t) \hat b_{in}(t')^\dagger \rangle = \delta(t - t')$. 

Using standard input-output theory \cite{Clerk2010}, we can calculate the output field in frequency space:
\begin{align}
    \hat a_{out}(\omega) &= \hat a_{in}(\omega) + \sqrt{\Gamma} \hat a_{0} (\omega) , 
\end{align}
If we define a quadrature $\hat q_{out}(\omega,\theta) = e^{i \theta} \hat a_{out}(\omega) + e^{-i \theta} \hat a_{out}^\dagger(-\omega)$, then we can define the squeezing spectrum as:
\begin{align}
    P(\omega) &= \min_{\theta \in [0,2\pi)} \langle \hat q_{out}(\omega,\theta) \hat q_{out}(-\omega,\theta) \rangle  , 
\end{align}
which gives us a measure of how squeezed the output field of the edge lattice site is versus frequency. This is depicted in \cref{fig:SSH_Output}, which shows narrow-band squeezed topological output light.

\subsection{Hofstadter Steady State Correlations}

\begin{figure}[H]
    \centering
    \includegraphics[width = 5in]{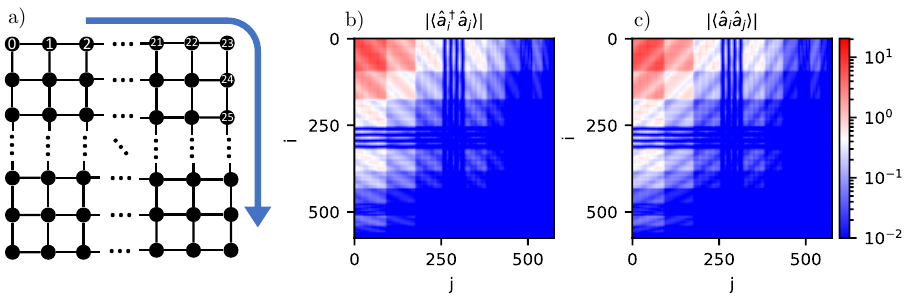}
    \caption{(a) shows the spiral ordering direction. (b) and (c) give the correlations $\langle \hat a_i^\dagger \hat a_j \rangle$ and $\langle \hat a_i \hat a_j \rangle$, respectively for a $24 \times 24$ site quarter flux Hofstadter lattice with $\1 = (11,23)$, $\0=(12,20)$, and $\eta = 0.999 \eta_c \sim 0.0007$, which is identical to the lattice in the main text.}
    \label{fig:Hofstadter_Full_Correlations}
\end{figure}

In this section, we provide all-to-all correlation plots for the Hofstadter lattices, as well as line cuts of the main text plots. To show all-to-all correlations, because the lattice lies on a 2D graph, it will be necessary to pick an ordering of the lattice sites. This is done most easily by choosing a ``spiral ordering'' where we start in the upper corner, and spiral around and into the lattice, see \cref{fig:Hofstadter_Full_Correlations}(a). By numbering the sites in this manner, the sites are grouped by their distance from the edge of the lattice, which allows one to easily discern patterns in the correlations, see \cref{fig:Hofstadter_Full_Correlations}(b-c). The upper-left block in both plots correspond to edge-edge correlation, which are exponentially enhanced. The correlations fall off exponentially as one moves into the bulk; i.e., down and to the right in these plots.

\begin{figure}[H]
    \centering
    \includegraphics[width=5in]{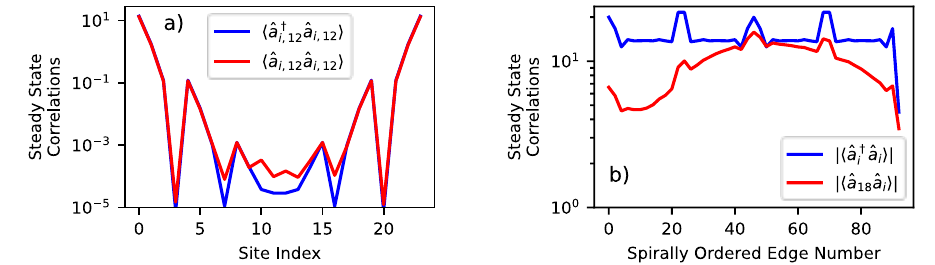}
    \caption{The lattice parameters are the same as in \cref{fig:Hofstadter_Full_Correlations}. a) shows a line cut through the center of the lattice, with the lattice $j$-coordinate fixed at 12. Note the exponential drop in correlations in the bulk compared to the edge. b) shows local photon density and correlations with the site $(18,23)$ (or just $18$ in the spiral ordering, see \cref{fig:Hofstadter_Full_Correlations}(a)) going around just the edge. Only every other lattice site is shown, as the correlations are identically zero on different sublattices. Note that the correlations are nearly uniform around the entire edge, except at the corners.}
    \label{fig:HofstadterLineCuts}
\end{figure}

Finally, we consider line cuts of the steady state correlations. First, we consider a horizontal line cut through the lattice, and look at local density and squeezing correlations, which shows exponential enhancement on the edges, \cref{fig:HofstadterLineCuts}(a). Secondly, in \cref{fig:HofstadterLineCuts}(b), we look at correlations just on the edge, with the upper-left hand corner marking site 0, and winding around in a clockwise pattern as is displayed in \cref{fig:Hofstadter_Full_Correlations}(a). This shows nearly spatially uniform correlations and densities around the edge, once taking into account that correlations between sublattices are identically zero.

\subsection{Hofstadter Entanglement Structure}
\begin{figure}[h]
    \centering
    \includegraphics[width = 5in]{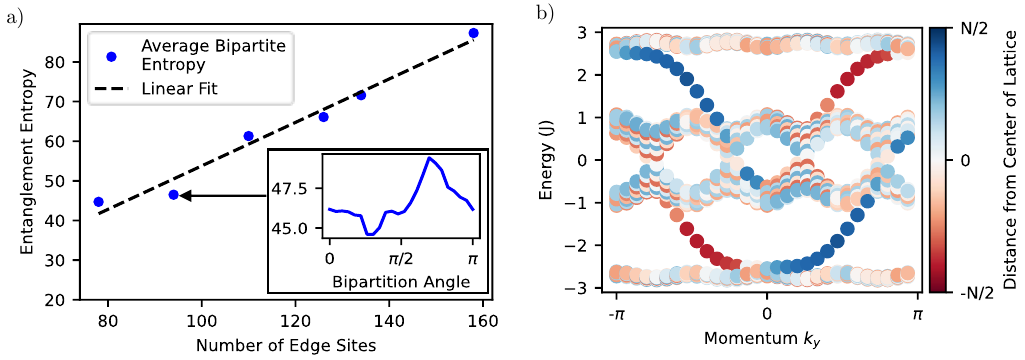}
    \caption{a) This depicts the bipartite entanglement entropy for a 20 $\times$ 20 to 50 $\times$ 50 quarter flux Hofstadter lattice. The dissipation is placed with site $\0 = (N/2 - 1,0)$ and $\1 = (N/2 + 1,2)$, on an $N \times N$ lattice. In all plots $\eta = .087 \sim 0.97 - 0.99 \ \eta_c$. Data points are shown only those for which the dimensionless pumping rate $\eta = 0.087 < \eta_c$, ensuring a steady state solution. Lattice bipartitions are made by defining a line through the center of the lattice at an angle $\theta$ to the horizontal, and data points are averaged over angles in the set $\{ \theta_i = \pi i /N | 0 \leq i < N\}$, see inset. The linear fit suggests the entanglement scales with a volume-law in the number of edge sites. Inset: Entanglement entropy versus bipartition angle for a $24 \times 24$ site lattice. This is representative for all $N$ shown, where the entropy varies by at most a few percent versus angle. b) shows the band diagram for a $30 \times 30$ quarter flux Hofstadter lattice, which is finite in $x$ and periodic in $y$, making $k_y$ a good quantum number. There are 4 distinct bands, the middle two touching at 4 Dirac points. The color shows average distance of a mode from the center of the lattice. The red (blue) band crossing edge modes are localized to the left (right) of the lattice.}
    \label{fig:HofstadterEntanglement}
\end{figure}

We can understand the entanglement structure of the edge modes in two ways: firstly, in real space, we observe that as $\eta \to \eta_c$, the real space entanglement approaches all-to-all correlations within each sublattice. For fixed $\eta$, the entanglement of the lattice averaged over all bipartitions scales linearly with the size of the edge. Tracing out the bulk modes to get a 1D edge picture, this constitutes a volume-law entanglement scaling, see \cref{fig:HofstadterEntanglement}(a). The inset shows an example of the entanglement entropy versus bipartition (characterized as the angle at which we cut the lattice), which is nearly bipartition independent.

As mentioned in the main text, we can also understand the entanglement structure from the band diagram, depicted in \cref{fig:HofstadterEntanglement}(b). Again assuming we can ignore the bulk modes and looking only at the edges, we will treat this as a 1D ring. Now, the real-space chiral symmetry multiplies a mode by a phase $e^{i \pi j}$ on site $j$, which is equivalent to sending the momentum $k \to k + \pi$. Further, we know that it sends the mode energy $\omega \to -\omega$. Thus, if we define the group velocity $\nu_g = \partial \omega/\partial_k$, it sends
\begin{align}
    \nu_g &= \frac{\partial \omega}{\partial k} \to \frac{\partial (-\omega)}{\partial (k + \pi)} = -\nu_g
\end{align}
Hence, the steady state pairs states with opposite group velocity, i.e.~chiral edge modes propagating clockwise with ones propagating counter-clockwise. In the band structure (see \cref{fig:HofstadterEntanglement}(b), assuming periodic boundary conditions in the $y$-direction), we can identify that modes localized to the left (red in the plot) pair with their partner at $k + \pi$ and negative energy. Similarly for right-localized sites in blue. 

\subsection{Topological Boundary Entanglement}
\begin{figure}[h]
    \centering
    \includegraphics[width = 1.7in]{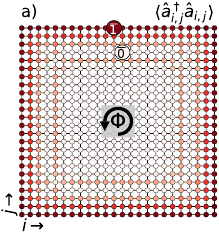}
    \includegraphics[width = 1.7in]{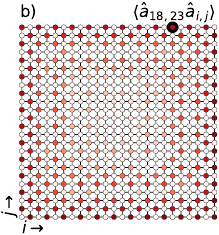}
    \includegraphics[width = 1.7in]{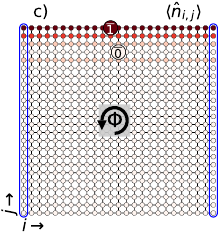}
    \includegraphics[width = 1.7in]{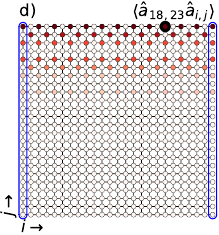}
    \caption{All four plots show steady state correlations for a 24 $\times$ 24 site quarter flux Hofstadter lattice. There is a single dissipator of the form $\hat L = \hat a_{\0} + \eta \hat a_{\1}^\dagger$, where the sites $\0 = (11,23)$ and $\1 = (12,20)$ are shown in a) and c). $\eta = 0.999 \eta_c$ in both plots. a) and c) show the local photon number density, and b) and d) show lattice correlations with the site (18,23), denoted by a dark circle. a) and b) correspond to a square lattice with open boundary conditions, giving a single extended edge. c) and d) have open boundary conditions in the $j$ direction, and are periodic in the $i$ direction; the lattice sites encircled in blue ellipses are identified with each other. Thus, the top-most and bottom-most rows of lattice sites are distinct edges separated by a 22-lattice site bulk. Note that the two edges are unentangled from each other completely, and the photon density is restricted to only a single edge in the steady state.}
    \label{fig:Hofstadter_PBC}
\end{figure}
Here, we prove the statement in the main text that in a topological lattice, two sites will have exponentially enhanced entanglement only if they are connected by a topological boundary. To show this, we assume that all of the bulk modes are nearly translationally invariant, and all of the edge modes have an exponentially decaying profile. We take the two lattice sites $\0, \1$ to have distances $x,y$ from the edge. Now, we can take $\eta \sim \exp(-\zeta_L(x - y))$, so that the delocalized bulk modes have an exponentially damped squeezing parameter, but the edge modes have $\tanh r_\alpha \sim 1$, and so they are approaching criticality. Clearly, if two sites exhibit some macroscopic entanglement, it must be via an edge mode, as the bulk modes are exponentially close to the trivial vacuum.

Now, let's assume that two sites are not connected by a topological boundary; that is to say, they live on different edges. There are two ways to entangle the two sites: firstly, one can imagine that there is a mode localized to one edge, whose minus energy partner is localized to another edge, and they then form a two mode squeezed vacuum, entangling the edges. However, this would violate our real space sublattice symmetry, since flipping a sign on every other lattice site cannot change which edge a mode is localized to. Thus, the only way to create some entanglement is if there is a mode that has a finite overlap with multiple edges. Let's define such a mode as $|\psi \rangle = (|L\rangle + |R\rangle)/\sqrt{2}$. Where $|L\rangle$ and $|R\rangle$ are localized to the left and right edges, respectively. Now, if we have a Hamiltonian $\H$, we know that
\begin{align}
    \H |\psi \rangle &= \epsilon |\psi \rangle \\
    \implies \H (|L \rangle + |R\rangle ) &= \epsilon  (|L \rangle + |R\rangle )
\end{align}
Invoking the assumption that the Hamiltonian is local, we observe that $\langle L | \H | R \rangle = O(e^{-\zeta_L N})$ is exponentially suppressed, where $\zeta_L$ is the localization length scale, and $N$ the distance between edges. Hence, in the limit $N \gg 1$, $|L\rangle$ and $|R\rangle$ are approximately degenerate eigenmodes. From here, we can break this down into two further cases: if $\epsilon = 0$ then we could have that the proper opposite energy pairs that respect the real space sublattice symmetry are linear combinations $|\psi_{\pm}\rangle = (|L\rangle \pm e^{i \phi} |R\rangle)/\sqrt{2}$, which is exactly the SSH model, and leads to no long range entanglement. 

The second option is that there are two other modes $|L'\rangle$ and $|R'\rangle$ with energy $-\epsilon$, and localized on the left and right, respectively. However, this just means that $|L\rangle$ is paired with $|L'\rangle$ and $|R\rangle$ is paired with $|R'\rangle$, again giving no cross-lattice entanglement.

This completes the proof that edge sites not connected by a topological boundary have exponentially suppressed correlations. A striking example of this is given in \cref{fig:Hofstadter_PBC}, which shows the correlations in a square Hofstadter lattice identical to \cref{fig:Hofstadter_Full_Correlations} and the main text, compared to the correlations where the boundary condition is made periodic along the horizontal direction, hence creating two distinct edges separated by a bulk. With open boundary conditions and a single topological edge, there is long range, volume law entanglement between all sides of the lattice (as shown in \cref{fig:HofstadterEntanglement}(a)). However, when there are two disconnected topological boundaries, there is no long range entanglement between the upper and lower edges, see \cref{fig:Hofstadter_PBC}(c-d).



\bibliography{ref}